# Small-size reflectionless band-pass filter

**Nickolay D. Malyutin** [1,*] **and Trinh To Thanh** [1,*]

[1] Tomsk State University of Control Systems and Radioelectronics (TUSUR), Tomsk, Russia
[*] E-mail : ndm@main.tusur.ru, thanhvodoi1995@gmail.com

**Abstract.** The paper presents a reflectionless stripline filter with reduced dimensions having a single bandwidth over a wide frequency range. The filter consists of coupled strip lines and RLC circuits included in the diagonal ports of the coupled strip lines. Calculations and measurements of the reflectionless stripline filter layout in the frequency range up to 4.8 GHz have been carried out. The obtained results show that the dimensions are reduced by a factor of two compared to previously developed reflectionless stripline filter designs. It is achieved by folding the coupled lines in the form of a meander formed by horizontally and vertically arranged conductors on horizontally and vertically oriented dielectric substrates. The proposed reflectionless stripline filter allows to solve the problem of improving the mass-dimensional parameters of the equipment and to provide signal selection with minimum reflection at out-of-band frequencies. It can be in demand in multichannel systems of wireless communications, radar, and measurements.

## 1. Introduction

Nowadays, reflectionless filters based on transmission lines and coupled lines find application in various radio-electronic equipment [1-3]. One of the main problems faced by designers and users of such filters is the limitations associated with their physical dimensions. Significant dimensions can have a significant impact on practical applications.

The paper [4] presents a description of a reflectionless band-pass filter. In the described design, the diagonal 3dB ports of the directional tap are connected to frequency-dependent loads that are reflective filters on coupled strip lines with resistive loads. However, both the directional tap itself and the frequency-dependent loads are distributed elements, which leads to an increase in filter size.

The research described in [5] is devoted to solving the problems of creating a reflectionless stripline filter. A detailed analysis of the influence of frequency characteristics of coupled stripline lines, which are loaded with concentrated RLC circuits on the characteristics of reflectionless stripline filter is presented. The use of concentrated elements partially solves the problem of providing an improvement in mass-dimensional performance. But this work uses the topology of transmission lines in the form of rectilinear segments, which have significant dimensions and limit the possibilities of miniaturization.

The main objective of this study is to develop the conductor topology of a reflectionless stripline filter with reduced dimensions by a factor of 2 while preserving the parameters of previously developed reflectionless stripline filters.

## 2. Design of reflectionless stripline filter

The design of the reflectionless stripline filter prototype [5] was implemented on a 60×24 mm substrate. The size of the reflectionless stripline filter can be reduced in different ways. One of them is to increase the relative dielectric constant of the substrates. However, as shown in [6], changing the substrates leads to inequality of phase velocities of in-phase and antiphase waves in coupled strip lines, which is accompanied by deterioration of frequency characteristics of the reflectionless stripline filter, in particular, the reflection coefficient at out-of-band frequencies increases. We applied a change in the topology of the conductors of coupled strip lines in the form of a meander. Also, as in [7-9], as coupled strip lines we used the coupled strip lines design, in which the coupled strips consist of two parts and have the shape shown on the cross section of the strip structure (figure 1).

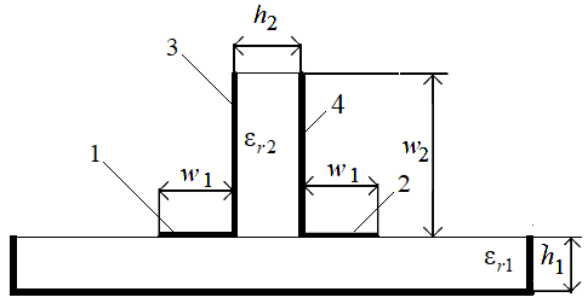

**Figure 1.** Cross section of strip coupled lines with vertically arranged substrate.

Horizontally located strips 1 and 2 are arranged on a substrate made of FR4 material with a thickness of $h_1 = 1,5$ mm and relative permittivity $\varepsilon_{r1} = 4,5$. The vertically arranged strips 3, 4 are on a vertically mounted substrate made of RO-3003 material with a thickness of $h_2 = 0.635$ mm and $\varepsilon_{r2} = 3,0$. The topology of horizontal conductors 1 and 2 coupled strip lines is shown in figure 2. They have the following parameters $w1 = 0.6$ mm, $w3 = 2.8$ mm, $l = 9$ mm, $h2 = 0.6$ mm, $t = 6$ mm.

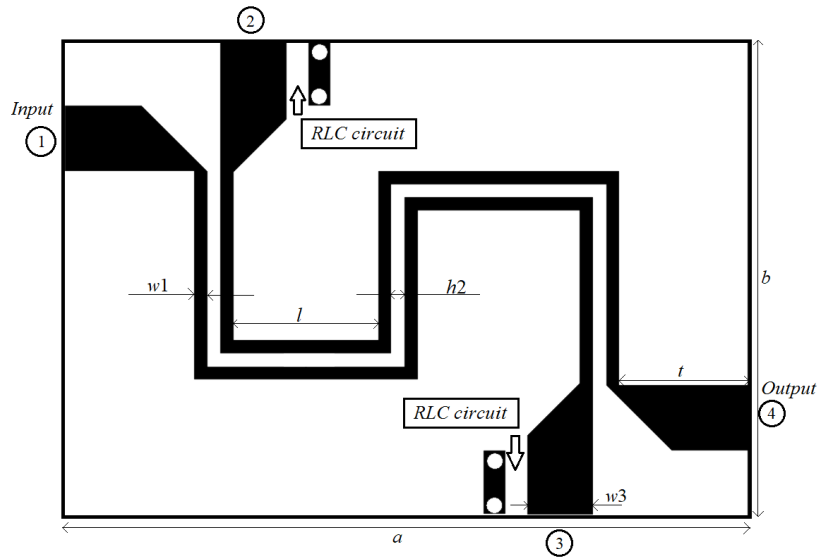

**Figure 2.** Topology of horizontal conductors.

Since the horizontal strips are convolved in a meander, the vertical strips were made on separate boards with the specified transverse dimensions according to figure 1, and the longitudinal dimensions correspond to the length of the parts of the horizontal conductors along the center line passing through the middle of the gap.

The equivalent circuit of the investigated reflectionless stripline filter is shown in figure 3. The circuit consists of five sections of coupled strip lines I-V, arranged perpendicularly to each other and two RLC circuits located in ports 2 and 3. Ports 1 and 4 respectively serve as the input and output of the reflectionless stripline filter. Each section is coupled strip lines with a longitudinal dimension of 9 mm. RLC-chain with the following parameters: $R0 = 50$ Ohm, $L0 = 23.1$ nH, $C0 = 1.2$ pF.

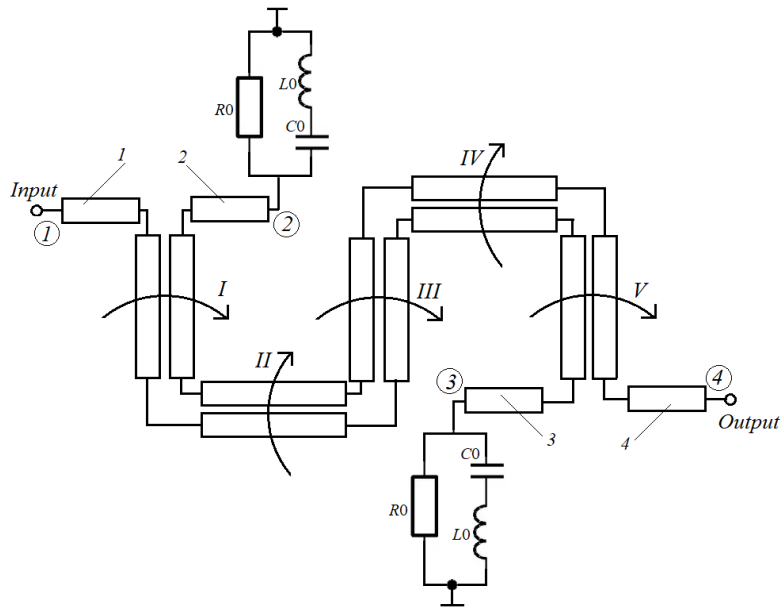

**Figure 3.** Equivalent circuit of a reflectionless stripline filter.

To ensure matching between input and output, it is necessary to use transmission lines of $w3 = 2.8$ mm width to achieve a wave impedance of 50 Ohm. Designing of the proposed reflectionless stripline filter at the first stage to simplify the calculation of *S*-parameters was carried out by replacing coupled strip lines in the form of a meander with a rectilinear equivalent line with a length of 45 mm. Subsequently, the calculation of parameters was carried out for a real design with reduced dimensions.

To calculate the primary parameters of coupled strip lines, the TALGAT software [10] was used. With the help of this software package the following primary parameters were obtained:

$$\mathbf{C} = \begin{bmatrix} 155.7 & -106.1 \\ -106.1 & 155.7 \end{bmatrix} \times 10^{-12}, \text{F/m - matrix of linear capacitances;}$$

$$\mathbf{L} = \begin{bmatrix} 3.857 & 2.633 \\ 2.633 & 3.857 \end{bmatrix} \times 10^{-7}, \text{H/m - matrix of linear inductances.}$$

Using the program [11], the frequency characteristics of the reflectionless stripline filter including reflection and transmission coefficients of the coupled strip lines were calculated. Figure 4 shows the plots of the dependence of S-parameters on frequency. The results show that the reflection coefficient remains at -20 dB. The center frequency is 0.94 GHz and the bandwidth is 0.2 GHz.

## 3. Experimental data

As a result of the development carried out, a reflectionless stripline filter was manufactured. Figure 5 shows the appearance of the layout, which has dimensions of 31×22 mm, which corresponds to a halving of the dimensions compared to the work [5]. Figure 6 shows the frequency characteristics of the filter. It can be seen from the graph that the filter functions at a center frequency of 0.94 GHz with a bandwidth of 1.8 GHz. At the same time, the reflection coefficient does not exceed -10 dB over the entire frequency range up to 4.8 GHz. The obtained experimental data confirm the possibility of creating a non-reflective band-pass filter with compact dimensions and characteristics close to the calculated ones.

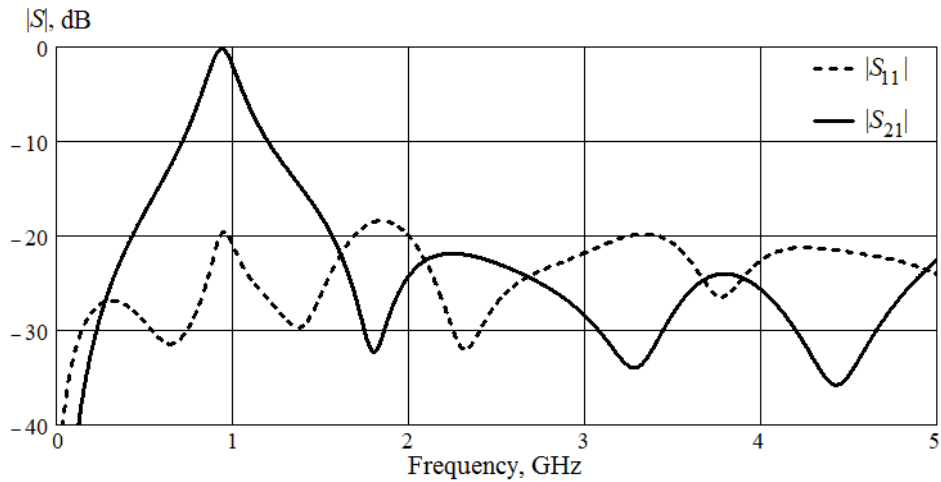

**Figure 4.** Frequency dependences of return loss and transmission coefficient of reflectionless stripline filter.

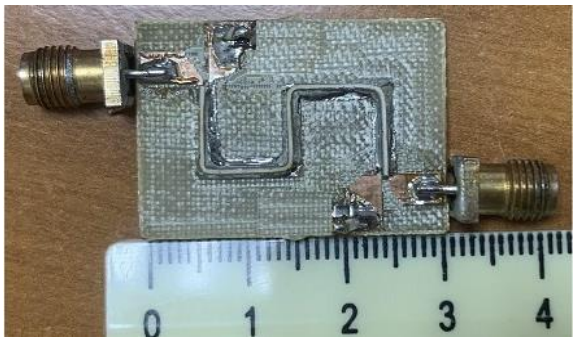

**Figure 5.** Manufactured model of a single-link reflectionless band-pass filter.

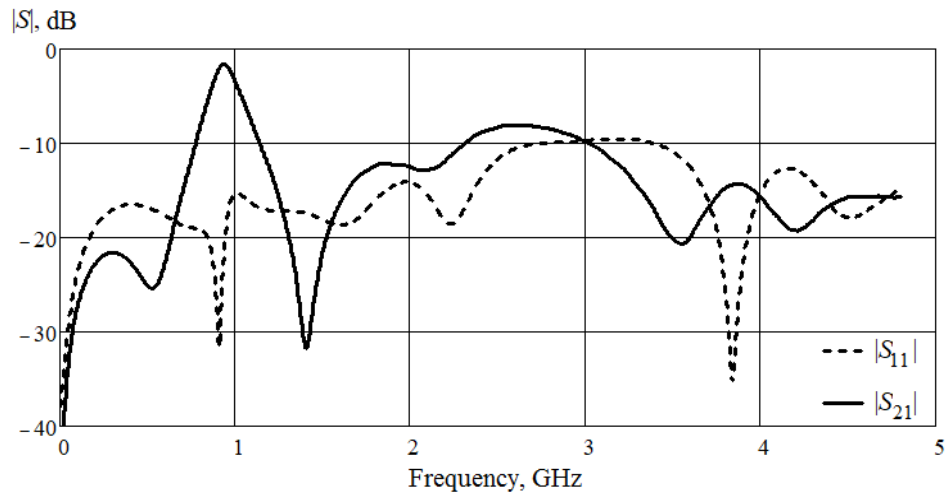

**Figure 6.** Experimental frequency dependences of return loss $|S_{11}|$ and transmission coefficient $|S_{21}|$ of the reflectionless stripline filter.

## 4. Conclusion

In the paper, a reflectionless stripline stripline filter has been presented with a reduction by a factor of two compared to previously developed reflectionless stripline filter designs. The filter provides operation at the center frequency of 0.94 GHz with a bandwidth of 6 GHz and a reflection coefficient that does not exceed -10 dB over the entire frequency range up to 4.8 GHz. Such single-link filters can be applied in radio systems where size reduction is required while minimizing reflection at out-of-band frequencies with non-repeating bandwidth. Cascading sufficiently well matched reflectionless stripline filter links will increase the selectivity, and the use of packing in pairs of links on the grounded base side will provide a negligible increase in the occupied volume of a two-link reflectionless stripline filter.

**Acknowledgments**
This publication was financially supported by the Ministry of Science and Education of the Russian Federation under Project No. FEWM-2023-0014.